\documentclass[preprint,12pt]{elsarticle}

\usepackage{amssymb}
\usepackage{amsmath}
\usepackage{subcaption}

\journal{Computers in Biology and Medicine}

\begin{document}

\begin{frontmatter}



\title{IARS SegNet: Interpretable Attention Residual Skip connection SegNet for melanoma segmentation}


\author[inst1]{Shankara Narayanan V}

\affiliation[inst1]{organization={Department of Computer Science and Engineering, Amrita School of Computing, Amrita Vishwa Vidyapeetham},
            city={Coimbatore},
            postcode={641112}, 
            country={India}}

\author[inst1,inst2]{Sikha OK}
\author[inst2,inst3]{Raul Benitez}

\affiliation[inst2]{organization={Department of Automatic Control, Universitat Politècnica de Catalunya},
            addressline={Av. d'Eduard Maristany}, 
            city={Barcelona},
            postcode={08019}, 
            country={Spain}}

\affiliation[inst3]{organization={Department of Automatic Control, Universitat Politècnica de Catalunya (BarcelonaTech)},
            city={Barcelona},
            postcode={08034}, 
            country={Spain}}
            
\begin{abstract}
Skin lesion segmentation plays a crucial role in the computer-aided diagnosis of melanoma. Deep Learning models have shown promise in accurately segmenting skin lesions, but their widespread adoption in real-life clinical settings is hindered by their inherent black-box nature. In domains as critical as healthcare, interpretability is not merely a feature but a fundamental requirement for model adoption. This paper proposes IARS SegNet an advanced segmentation framework built upon the SegNet baseline model. Our approach incorporates three critical components: Skip connections, residual convolutions, and a global attention mechanism onto the baseline Segnet architecture. These elements play a pivotal role in accentuating the significance of clinically relevant regions, particularly the contours of skin lesions. The inclusion of skip connections enhances the model's capacity to learn intricate contour details, while the use of residual convolutions allows for the construction of a deeper model while preserving essential image features. The global attention mechanism further contributes by extracting refined feature maps from each convolutional and deconvolutional block, thereby elevating the model's interpretability.  This enhancement highlights critical regions, fosters better understanding, and leads to more accurate skin lesion segmentation for melanoma diagnosis. This study primarily focuses on the interpretation of performance improvements in the base model resulting from the integration of each of these three components. To comprehensively assess the performance gain achieved with each addition, we employ two sets of evaluation metrics, quantifying performance based on both regions and contours. The results underscore the superior segmentation capabilities of the proposed architecture compared to the SegNet and U-Net models. Notably, it provides interpretable results, particularly when applied to the PH2 dataset. 
\end{abstract}



\end{frontmatter}


\section{Introduction}
\label{sec:introduction}
Melanoma, the most fatal variant of skin cancer, originates from melanoc-\\ytes responsible for producing melanin \cite{dildar2021skin}. Despite representing only 1\% of reported skin cancer cases, melanoma contributes to a staggering 80\% of skin-cancer-related deaths. The alarming rise of melanoma in predominantly fair-skinned countries over the past decade has made this the 5th most common cancer diagnosed in the United States of America. While other cancer types are expected to decrease or stabilize, skin cancer, especially melanoma, poses a severe and growing threat \cite{matthews2017epidemiology}. Early detection of melanoma is crucial, as it becomes progressively more challenging to treat in advanced stages. Traditionally, doctors rely on the biopsy method for skin cancer detection, involving the removal of a sample from a suspected skin lesion for examination to determine its cancerous nature. However, this procedure can be painful, slow, and time-consuming \cite{dildar2021skin}. With the rapid advancement of technology, Computer-Aided Diagnosis (CAD) has emerged as a promising approach for screening and early detection of melanoma \cite{masood2013computer}. The typical automated skin cancer detection pipeline involves acquiring the image, preprocessing it, segmenting the preprocessed image, extracting relevant features, and classifying the lesion \cite{okur2018survey}. Over the past decade, recent advancements in deep learning techniques have significantly contributed to the effective detection and diagnosis of melanoma \cite{dildar2021skin, naeem2020malignant, pacheco2019recent}. These sophisticated approaches have demonstrated promising results in accurately identifying and distinguishing malignant skin lesions, aiding medical professionals in making timely and precise diagnostic decisions. Combining image analysis, deep learning algorithms, and computational power has opened new horizons in skin cancer detection, offering improved efficiency and reliability for early diagnosis and optimal treatment outcomes. 
Precise skin lesion segmentation is pivotal in elevating the accuracy and dependability of subsequent lesion classification. Through meticulous delineation of lesion boundaries, segmentation becomes a vital factor in substantially augmenting the precision of subsequent classification algorithms \cite{arora2021automated}. This pivotal stage within the diagnostic process holds the potential to propel the field of skin cancer detection forward. It offers more resilient and trustworthy results, ultimately leading to enhanced patient care and treatment \cite{navarro2018accurate}.
Deep-learning segmentation models, such as U-Net \cite{ronneberger2015u} and SegNet \cite{ninh2019skin}, have demonstrated encouraging performance in skin lesion segmentation. However, their complex black-box architecture restricts their usability in the segmentation process for expert clinicians. In high-stakes tasks such as skin cancer diagnosis, interpretability emerges as a vital aspect to facilitate cross-verification by human experts \cite{rezk2023interpretable}.

In this paper, we have taken significant strides toward achieving a more interpretable, accurate, and trustworthy deep-learning segmentation model. Our approach involves the integration of various computational modules into the state-of-the-art SegNet architecture, resulting in a novel and improved model. By meticulously examining the contributions of each component within the network, we have succeeded in enhancing its transparency and efficiency. Furthermore, the feature maps extracted from the encoding and decoding blocks of the segmentation model play a pivotal role in validating the final segmentation process. These feature maps provide valuable evidence and essential information, enabling human experts to make more informed inferences. With interpretability at the forefront of our approach, we aim to bridge the gap between powerful deep-learning algorithms and the need for human expertise in verifying critical diagnoses.

Our work represents an important step towards building more transparent and reliable systems for skin lesion segmentation, ultimately supporting medical professionals in making well-informed decisions for improved patient care. The major contributions of the proposed works are as follows:
\begin{enumerate}
    \item A  self-interpretable segmentation network with residual convolutions and attention mechanism to achieve a higher segmentation accuracy and a better definition of the lesion contour. 
    \item The interpretability of residual convolutions and attention mechanisms adds to the transparency of the segmentation model, making it a valuable tool for reliable and precise skin lesion segmentation tasks.
    \item Extensive quantitative and qualitative validation of the effectiveness of the proposed segmentation architecture on the PH2 dermoscopic dataset. To quantitatively evaluate the accuracy of the segmented contours, we use different metrics to quantify the overall accuracy of the segmentation and to evaluate the detection of the lesion's contour. This comprehensive validation process ensures a thorough understanding of the architecture's performance, providing valuable insights into its segmentation capabilities. 
\end{enumerate}

\section{Related Works}\label{sec:related_works}
Recently, many developments have been made in solving the skin lesion segmentation task. Although the skin lesion segmentation task has been heavily researched, the task is far from being fully solved due to the complexity of the dermoscopic lesion images \cite{szegedy2015going}. This section showcases various noteworthy developments closely allied to our work. The remainder of this section will be divided into three parts, discussing the traditional methods, the deep learning techniques, and other attention-based models for skin lesion segmentation.  
\subsection*{Traditional Skin Lesion Segmentation Techniques}
Researchers have developed models using thresholding algorithms combined with clustering \cite{emre2013lesion} for skin lesion segmentation. Kajsa Møllersen \textit{et al.} \cite{mollersen2010unsupervised} describes a threshold technique after density analysis and \cite{yueksel2009accurate} talks about a fuzzy logic-based automatic thresholding algorithm. Several edge-based and region-based segmentation techniques were also presented \cite{abbas2011lesion} \cite{emre2008border}  to obtain the fine borders from the lesion image. The histogram-based clustering methods proposed in \cite{ashour2018hybrid} help differentiate various affected parts in a lesion based on color details. The common trait between all the above-mentioned traditional methods is their dependence on intensity-based features. Precisely due to this reason, these methods are not capable of understanding the contextual information of the lesion, and that is where deep learning methods have an advantage. 

\subsection*{Deep Learning Based Skin Lesion Segmentation Techniques}
At the outset, Deep Learning found its primary application in skin lesion classification \cite{lopez2017skin}. The authors employed transfer learning with VGGNet to classify skin lesions, leveraging the ISIC dataset \cite{gutman2016skin}. 
The results showed that their Deep Learning approach achieved higher accuracy and AUC-ROC scores than traditional Machine Learning methods, which relied on hand-crafted features.

Rasel \textit{et al.} \cite{rasel2022convolutional} exhibited the use of CNNs in skin lesion segmentation and compared multiple configurations based on the activation functions. Even though Deep Learning solutions do not require pre or post-processing steps, there is a need to find the correct set of hyperparameters. Rasel \textit{et al.} experimented on various combinations of the parameters such as stride, dilation factor, max epochs, convolutional filter, and max-pooling filter. The minimum number of training images required to achieve a significant result was also explored. Hasan \textit{et al.} \cite{hasan2020dsnet} proposed a Dermoscopic Skin Network (DSNet) by using depth-wise separable convolutions in place of the standard convolutions. This improvement led to better performance than well-established networks such as U-Net \cite{ronneberger2015u} while having a reduced number of parameters. Similarly, DeepLabv3+ \cite{chen2018encoder} is a Fully Convolutional Network (FCN) that incorporates atrous convolutions, which enable the network to efficiently compute dense feature maps in parallel, all without a significant increase in the number of parameters. The model uses an encoder-decoder architecture where the encoder part encodes multi-scale contextual information by applying atrous convolutions at multiple scales. In contrast, the decoder part refines the segmentation results along the object boundaries. 
Goyal \textit{et al.} \cite{goyal2019skin} enhanced DeepLabv3+ by proposing an ensemble architecture for skin lesion segmentation. The architecture has a preprocessing step, inferencing from DeepLabV3+ \cite{chen2018encoder} and Mask R-CNN \cite{he2017mask}. Finally, a post-processing step was applied to the output image of the DeepLabV3+ model involving basic morphological operations such as opening and closing to remove artifacts accrued during segmentation. Kumar \textit{et al.} \cite{kumar2018u} proposed U-SegNet, a hybrid of both the SegNet and the U-Net architectures. This architecture uses SegNet as the base and includes a skip connection (as present in U-Net) at the uppermost layer to incorporate feature maps with fine details. Capturing this multiscale information enhances the performance of the model with a minor trade-off in terms of an increase in the number of parameters compared to the original SegNet. Şaban \textit{et al.} \cite{ozturk2020skin} proposed an improved Fully Convolutional Network (iFCN) architecture to segment full-resolution skin lesion images without employing any pre or post-processing steps. This architecture includes residual connections that allow the network to learn residual features. These residual connections help improve the accuracy of the network by allowing the model to learn more complex features. Due to these connections, the iFCN architecture can better capture the lesion edges' details better and improve the segmentation accuracy. Most importantly, despite all these improvements, using residual connections helps reduce the number of parameters in the network. A reduced number of parameters is better for two main reasons. The model is less likely to overfit, and as there are fewer parameters, the model is computationally more efficient. 

\subsection*{Attention-based Skin Lesion Segmentation Techniques}
Attention mechanisms are useful for segmentation tasks in many different ways. The attention mechanism can preserve the two-dimensional structural information present in images and improve segmentation accuracy. Liu  \textit{et al.} \cite{liu2023van} introduced an efficient skin lesion segmentation approach with a multi-scale cross-attention mechanism, an enhanced version of SENet \cite{hu2018squeeze}. This mechanism includes two key components: the multi-scale channel attention (MSC-attention) block and the cross-scale feature fusion (CSFF) block. The MSC-attention block has global and local attention modules, capturing both global and local channel dependencies. The CSFF block incorporates up-sampling and feature fusion modules to create a comprehensive representation of the input image.
Tran  \textit{et al.} \cite{tran2022fully} introduced an efficient skin lesion segmentation architecture that employs additive attention mechanism to emphasize relevant features while suppressing irrelevant ones. Additionally, they integrated fuzzy logic to account for uncertainty and imprecision in the segmentation process. The segmentation is guided by the fuzzy energy-based shape distance as the loss function, computed using attention maps generated by the attention gate. These attention maps indicate the relevance of feature maps for segmentation, and the fuzzy energy-based distance measures the similarity between the segmentation boundary and attention feature maps.

\section{Proposed Segmentation Model}\label{sec:proposed_segmentation_model}
This section describes the proposed segmentation model in detail. The proposed architecture for efficient segmentation of skin lesions is built with SegNet as the baseline model.  Initially designed for road scene segmentation, SegNet lacked essential features required for precise medical image segmentation tasks. The SegNet model prioritized memory efficiency, and real-time video feed processing, often at the expense of segmentation accuracy. As a result, certain compromises were made in the decoder's upsampling techniques, which led to a reduction in the model's segmentation accuracy. In SegNet, pooling indices generate sparse upsampled maps, which are then convolved with trainable filters in the decoder blocks, leading to reduced computational speed but compromised segmentation accuracy. In the context of skin lesion segmentation, where precise delineation of clinically significant features such as boundaries is of utmost importance, the trade-off between speed and accuracy becomes untenable. The major drawbacks of the SegNet model for skin lesion segmentation include 1) Lack of precise upsampling techniques. 2) Lack of an attention mechanism. To overcome these constraints, the proposed segmentation model introduces three significant enhancements to the existing SegNet architecture:
\begin{enumerate}
    \item[a.] Incorporation of skip connections, enabling precise upsampling and better feature reconstruction.
    \item[b.] Introduction of residual convolution blocks to enhance information processing within the encoding and decoding components of the model.
    \item[c.] Integration of an attention mechanism to facilitate both efficient and accurate localization of the region of interest (ROI).
\end{enumerate}
 Furthermore, the modifications were done in a completely interpretable manner, with extensive visualizations and performance measures that justified the addition of every component to the architecture. This facilitates the extraction of human-understandable feature maps that provide an understanding of how the model learns generalized features from an input image.

\begin{figure}[ht]
    \centering
    \includegraphics[width=1\textwidth]{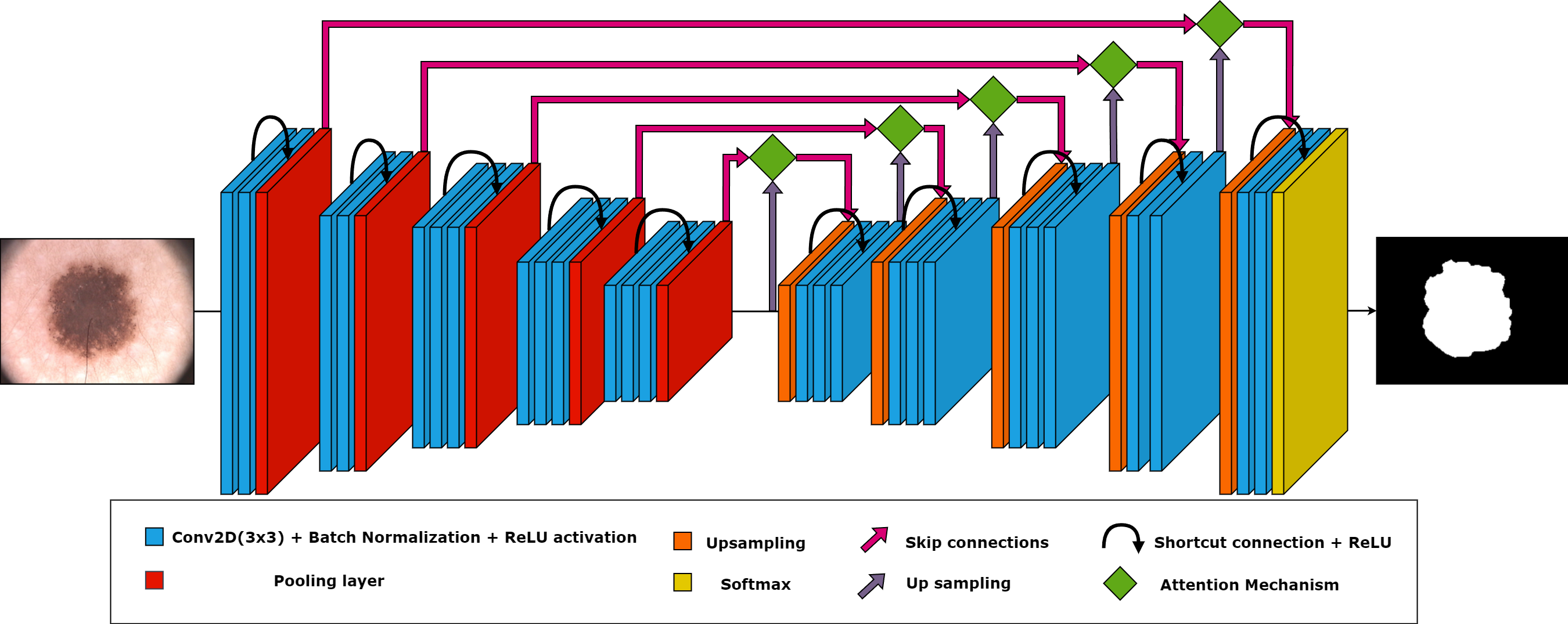}
    \caption{Proposed IARS SegNet segmentation framework}
    \label{prop_arch}
\end{figure}

\begin{enumerate}
    \item[a.] \textbf{Skip connections:} Skip connections facilitate the transfer of feature maps from encoding (down-sampling) layers to the corresponding decoding (up-sampling) path. This enables the preservation of coarser and finer details in the final segmentation map, thus enhancing the model's ability to retain critical spatial information. The inclusion of skip connections to the proposed segmentation model is inspired by the U-Net architecture. U-Net is renowned for its exceptional multiscale information capture, facilitated by the presence of skip connections. On the other hand, SegNet excels in faster processing and reduced parameter requirements by passing pooling indices to the upsampling layers. By incorporating skip connections into the SegNet architecture, we enable the model to leverage the benefits of both approaches. This integration enhances the SegNet's ability to capture multiscale information and finer details, addressing the limitation it had in this regard. Moreover, to manage the increased number of trainable parameters resulting from skip connections, we utilize $1 \times 1$ convolutional layers similar to the implementation found in GoogLeNet \cite{szegedy2015going}. This strategic implementation allows us to capture finer and coarser details without significantly increasing the overall parameter count, effectively optimizing the model's performance. Also, it highlights and weighs the contours of the skin lesion better than before the inclusion. The feature maps extracted from the decoder blocks are more visually interpretable, as the shape and border details are preserved through the skip connections. Furthermore, the model after the inclusion of the skip connections does not have any fully connected layers as it uses only the valid part of the convolutions, considerably reducing the trainable parameters. 

A U-Net-style skip connection is used in place of the pooling indices. This provides retention of relevant granular contextual information present in the original image. This is an essential modification mainly because this facilitates the propagation of the borders of the skin lesion, enabling accurate representation in the final segmentation map. This is verified by a visual interpretation of the feature maps generated from the corresponding decoder blocks before and after adding the skip connections. The inclusion of the skip connections highlights and weighs the contours of the skin lesion. Also, the feature maps extracted from the decoder blocks are more visually interpretable as the shape and border details are preserved through the skip connections. Furthermore, the model after the inclusion of the skip connections does not have any fully connected layers, as it uses only the valid part of the convolutions \cite{ronneberger2015u}.

 \item[b.] \textbf{Residual convolutions:}  Inspired by residual learning \cite{he2016deep}, our architecture adopts residual convolutions to promote deeper network training and alleviate the vanishing gradient problem. This fosters the efficient propagation of gradients and enables the model to capture more intricate patterns within the data. The residual convolutions replaced the conventional convolutions to improve the precision of the upsampling mechanism. The model can be trained to a greater depth by implementing dense residual connections between layers. Including dense residual connections enables the preservation and smooth propagation of a fine-tuned signal throughout the network. Upon passing the unchanged input to the residual blocks (as shown in Figure \ref{res_conv}), the process of preserving relevant information becomes considerably more efficient. This approach mitigates the vanishing gradient problem and facilitates the training of a deeper and more effective model. The utilization of dense residual connections in the proposed architecture significantly contributes to its improved performance and capability for accurate skin lesion segmentation. After the inclusion of the residual convolutions, the proposed network has the same number of network parameters compared to that of the popular U-Net model for biomedical image segmentation. The shortcut connections perform identity mapping, and their outputs are added to the outputs of the stacked layers. This ensures that the model's complexity remains the same while significantly boosting segmentation accuracy.
\begin{figure}[h]
    \centering
    \includegraphics[width=\textwidth]{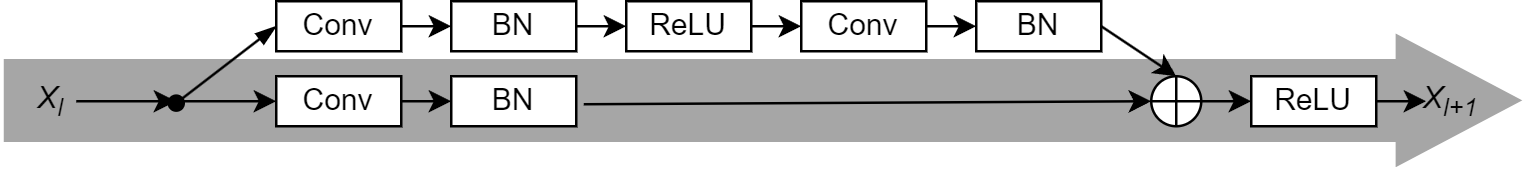}
    \caption{Residual Convolution (RC) component}
    \label{res_conv}
\end{figure}

The conventional convolutions were replaced by the residual convolutions to improve the precision of the upsampling mechanism. Apart from that, there are several advantages of including residual convolutions. Mainly, it eases the training of deep architectures, and the feature accumulation ensures better feature representation for the segmentation task. In the proposed network, after the inclusion of the residual convolutions, the number of network parameters does not change as the shortcut connections perform identity mapping, and their outputs are added to the output of the stacked layers \cite{he2016deep}. Because of this, the complexity of the model remains the same while improving the accuracy of the segmentation.

\item[c.] \textbf{Attention Mechanism:} To focus on relevant regions and suppress noise-inducing elements, attention mechanisms are introduced. By dynamically weighting the feature maps, the model can emphasize informative regions while reducing the impact of less relevant areas, improving segmentation accuracy. An attention mechanism was integrated into the network to enhance performance and reduce the number of False Positives (FP) in the final segmentation map. As shown in Figure \ref{attn_mech}, the attention mechanism is a 2D variant of the attention gate proposed in \cite{oktay2018attention}. The attention mechanism functions progressively, efficiently suppressing feature responses in irrelevant background regions without cropping the Region Of Interest (ROI) as in hard-attention mechanisms \cite{oktay2018attention}. By dynamically weighting the feature responses, the attention mechanism allows the network to focus on the most relevant regions while excluding irrelevant background information. This targeted attention helps to refine the segmentation process, leading to improved accuracy and a reduction in False Positives, ultimately bolstering the network's performance for skin lesion segmentation.
     \begin{figure}[h]
    \centering
    \includegraphics[width=\textwidth]{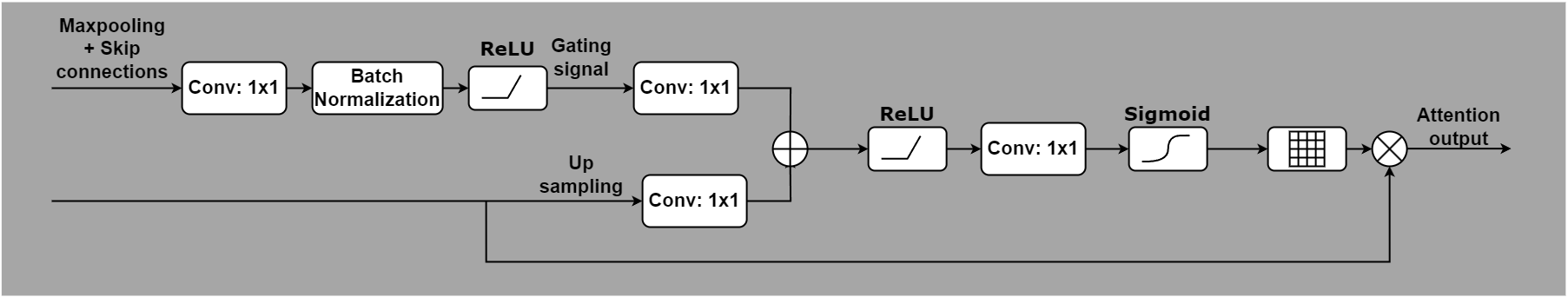}
    \caption{Attention Mechanism (AM) component}
    \label{attn_mech}
\end{figure}
Using a cascade model for extracting features will prove to be computationally very ineffective, as there is a lot of repetition of the low-level features. The effect of the attention mechanism is visible in the extracted Maximum Intensity Projection maps (Figure \ref{MIP}) from the decoder blocks. The features maps show how the model increasingly focuses on various parts of the input lesion image and finally converges on the lesion. Hence, the addition of attention mechanism improves the interpretability of the feature maps extracted from the encoder and decoder blocks, giving insight into the segmentation process of the entire architecture.
\end{enumerate}
Collectively, these three enhancements empower the presented architecture to effectively overcome the limitations of the original SegNet for the skin lesion segmentation task. As a result, we obtain a segmentation model that is more resilient and efficient, with the capability to accurately capture both global and local features within the input data. Figure \ref{prop_arch} illustrates the proposed enhanced segmentation architecture.

\section{Dataset}\label{sec:dataset}
The PH2 \cite{mendoncca2015ph2} dermoscopic dataset containing 200 images was used to train and test the proposed architecture. The original image resolution of 768x500 was downsized to 192x256. This resizing curtailed the count of trainable parameters and expedited the training process. To augment the input images, we employed Keras's ImageDataGenerator, configured to introduce random flipping and rotations during the training phase. Rotational transformations ranged from -40 to 40 degrees, and horizontal flipping was applied. By seamlessly integrating these stochastic transformations into the training process, our model gained the capability to recognize patterns even in the presence of such variations.
\begin{figure}[ht]
\centering
\includegraphics[width = \textwidth]{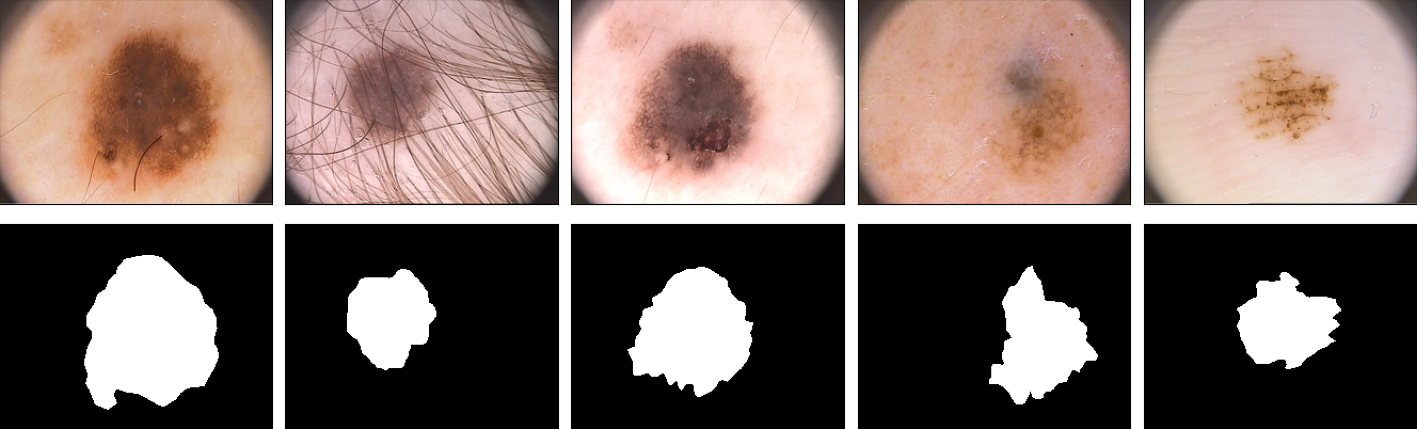}
\caption{Sample images from The PH2 dataset. The first row displays all the lesion images and the second row shows the corresponding ground truth segmentation masks.}
\label{lesions}
\end{figure}
Figure \ref{lesions} shows sample skin lesion images and their corresponding ground truth segmentation masks from the PH2 dataset. It is evident from the figure that these images encompass extraneous elements such as hair, oil bubbles, etc. These undesired components pose a challenge to the accurate segmentation of lesions. So, it becomes imperative to implement the right strategies to address and mitigate the impact of these elements on segmentation accuracy. 

\section{Evaluation Metrics}\label{sec:evaluation_metrics}
This section describes the assessment metrics employed for evaluating the proposed model. The augmentative elements integrated into the model contribute progressively to enhancing segmentation accuracy. Moreover, these components are strategically incorporated to enhance border precision by introducing skip connections and residual convolutions. An attention mechanism is seamlessly integrated into the architecture to refine lesion localization further. Two distinct sets of metrics encompass the performance evaluation. These metrics collectively encapsulate both the accuracy of segmentation and the intricacies of contour shape. Specifically, they provide a comprehensive evaluation framework that captures the nuanced aspects of segmentation quality and contour fidelity.
\begin{enumerate}
\item[a.] Region-based metrics (quantify segmentation): IoU, TNR, FNR, TPR, FPR, and Dice score
\item[b.] Contour-based metrics (quantify contour details): Elliptical Fourier Descriptors (EFDs), Hu Moments 
\end{enumerate}
This distinction between the metrics highlights the model's improvement in general segmentation and the improvements in the contours captured. While the improvement in contours is a subset of the broader segmentation accuracy improvement achieved by the architecture, the clinical significance of contours underscores the importance of this study.  In the following subsections, we delve deeper into the interpretability of the model by dissecting the contributions of its individual components, further enhancing our understanding of how each element contributes to the overall performance.

\subsection*{Region-based metrics: Segmentation performance} 
Semantic segmentation is a task where each pixel is assigned a class in the final segmentation map. One way of measuring the model's accuracy is pixel-wise accuracy, but in the case of class imbalance, this measure does not represent the true effectiveness of the model. There are several examples in the dataset where the lesion size is too small, and the prediction segmentation map can become immune to class imbalance as there are more background pixels (black pixels) than the actual segmented lesion (white pixels). IoU is defined to measure the overlap between the predicted segmentation map ($g'$) and the ground truth segmentation map ($g$) and is defined as follows,

    \begin{equation}
    IoU(g, g') = \frac{|g \cap g'|}{|g \cup g'|}
    \label{eq:iou}
    \end{equation}

    The numerator in the equation \ref{eq:iou} corresponds to the regions of overlap between the ground truth segmentation mask and the predicted segmentation mask. The denominator is the combination of both masks. 
    
    Based on the number of correctly and wrongly classified pixels, we also consider four metrics in our study:  True Positives (TP) define the number of pixels correctly classified to the positive (foreground) class;  True Negatives(TN) define the number of pixels correctly classified to the negative (background) class; False Positives (FP)  define  the number of pixels incorrectly classified as the positive (foreground) class and False Negatives (FN)  define the number of pixels incorrectly classified as the negative (background) class. Based on these four measures, the True Positive Rate (TPR), True Negative Rate (TNR), False Positive Rate (FPR), and False Negative Rate (FNR) can be calculated as in equation.\ref{eq:tpr}, \ref{eq:tnr}, \ref{eq:fpr} and \ref{eq:fnr} respectively.

    \begin{equation}
        \text{True Positive Rate (TPR)} = \frac{\text{TP}}{\text{TP} + \text{FN}}
        \label{eq:tpr}
    \end{equation}
    
    \begin{equation}
        \text{True Negative Rate (TNR)} = \frac{\text{TN}}{\text{TN} + \text{FP}}
        \label{eq:tnr}
    \end{equation}
    
    \begin{equation}
        \text{False Positive Rate (FPR)} = \frac{\text{FP}}{\text{FP} + \text{TN}} 
        \label{eq:fpr}
    \end{equation}

    \begin{equation}
        \text{False Negative Rate (FNR)} = \frac{\text{FN}}{\text{FN} + \text{TP}}
        \label{eq:fnr}
    \end{equation}
    Apart from the above-mentioned measures, we also consider Dice Score, which is the ratio of twice the area of overlap between the ground truth segmentation mask and the predicted segmentation mask to the sum of the areas of both segmentation masks as in the equation.\ref{eq:dice}.
    \begin{equation}
            Dice(g, g') = \frac{2*|g \cap g'|}{|g| + |g'|}
            \label{eq:dice}
    \end{equation}
    \subsection*{Contour-based metrics: Capturing contour’s shape} 
    There are several characteristic features present on the skin lesion, which indicates its malignity. One such important clinically relevant feature is the inconsistent pigment pattern on the borders of the lesion \cite{nachbar1994abcd}. This is captured effectively by the contours of the segmentation map. A good representation of the lesion's borders plays a crucial role for a better CAD diagnosis. To establish the efficiency of the proposed model in extracting the borders of a skin lesion, we are using Elliptical Fourier Descriptors (EFDs) \cite{kuhl1982elliptic} and Hu Moments \cite{huang2010analysis}.
    
    i) \textit{Elliptical Fourier Descriptors}: The contours of the ground truth segmentation masks and the predicted segmentation masks are extracted using the chain code boundary descriptor. The extracted contours are used for calculating the individual contours' Elliptical Fourier Descriptors (EFDs). We then use a Python package called PyEFD \cite{pyefd} to extract the Fourier coefficients $A_n$, $B_n$, $C_n$, and $D_n$ by passing a closed contour and the number of harmonics. The returned Fourier coefficients are normalized; they are rotational and size invariant. The output vector will be of the shape $(n, 4)$ where $n$ is the total number of harmonics chosen, with four coefficients per harmonic. 
    
    \begin{equation}
        x_{n} = \sum_{n=1}^{N} A_{n} \cos(nt) + B_{n} \sin(nt)
        \label{eq:hu_x}
    \end{equation}
    \begin{equation}
        y_{n} = \sum_{n=1}^{N} C_{n} \cos(nt) + D_{n} \sin(nt)
        \label{eq:hu_y}
    \end{equation}
    \begin{align*}
            N & : \text{Maximum number of harmonic amplitudes used in the construction} \\
            n & : \text{Harmonic amplitude index} \\
            t & : \text{Evaluation angle}
    \end{align*}
    Fourier Coefficients are then used to reconstruct the contours. The coordinates for reconstructing the contour are calculated using equations \ref{eq:hu_x} and \ref{eq:hu_y}. The overlap between the actual and reconstructed contours relies on the number of harmonics chosen. For simpler shapes, fewer harmonics are enough to describe the contour accurately. Higher-order harmonics (large $n$) better reproduce all the finer details in the contour and are used for complex shapes.  Choosing a lower-order harmonic will give a lesser error for simpler shapes but perform poorly for complex shapes. On the contrary, taking a larger number of harmonics for all shapes will be computationally inefficient. Hence, there is a need to get an optimal number of harmonics for representing all the contours, and this was found empirically by analyzing the error rate for each harmonic on the entire dataset. It was found that an optimal number of harmonics for the PH2 dataset would be $100$; thus, each contour will return an EFD coefficient vector of dimension $(100,4)$.

    The Fourier Coefficients of the ground truth contours and the predicted segmentation contour are used to carry out a statistical multivariate analysis. The Fourier Coefficient vectors are compared using the Mahalanobis distance (equation \ref{eq:mahdist})  to analyze the similarity of contour representation. The model with a lower mean Mahalanobis distance gives the best contour representation. The statistical significance of the distribution of the Mahalanobis distances between the base SegNet model and the final enhanced model is statistically tested using a non-parametric Wilcoxon rank-sum test. 
    Let, $e$ and $\hat{e}$ be the ground truth and predicted mask Fourier Coefficient vectors and $\Sigma$ the sample covariance matrix of the distribution. We define a similarity metrics using the Mahalanobis distance between the two distributions $e$ and $\hat{e}$ 
    \begin{equation}
        D_{\text{M}}(e,\hat{e}) = \sqrt{(e - \hat{e})^T \Sigma^{-1} (e - \hat{e})}
        \label{eq:mahdist}
    \end{equation}
    
    ii) {\em Hu Moments}: The second measure for analyzing the contours of the segmentation mask is by using the Hu Moments \cite{huang2010analysis}. A total of seven Hu Moments are calculated using central moments that are invariant to image transformations. The first six moments are invariant to translation, scale, rotation, and reflection, while the $7^{th}$ moment will change its sign for image reflection. Two vectors will store the seven Hu Moments for the prediction and ground truth segmentation maps. A log transform is applied to all the moments to make them comparable in scale. For each image, there are seven moments invariant of translation, rotation, and scale, describing its shape. The similarity between the contours of the ground truth and the segmentation mask is then calculated using Euclidean distance measure as shown in equation \ref{eq:similarity}. Where $\Phi$ denote the Hu Moments of the ground truth segmentation mask and $\hat{\Phi}$ denote the Hu Moments of the predicted segmentation mask:
    \begin{equation}
    Euclidean Distance(\Phi,\hat{\Phi}) = \sqrt{\sum_{i=1}^{7} (\Phi_i - \hat{\Phi}_i)^2}, 
    \label{eq:similarity}
\end{equation}
where $\Phi = [\phi_{1} \, \phi_{2} \, \phi_{3} \, \phi_{4} \, \phi_{5} \, \phi_{6} \, \phi_{7}]^T$ and $\hat{\Phi} = [\hat{\phi}_{1} \, \hat{\phi}_{2} \, \hat{\phi}_{3} \, \hat{\phi}_{4} \, \hat{\phi}_{5} \, \hat{\phi}_{6} \, \hat{\phi}_{7}]^T$. 
A detailed mathematical formulation of Hu Moments is provided in Appendix A.
\subsection{Experiments}\label{sec:experiments}
The proposed architecture for efficient semantic segmentation of skin lesions is built with SegNet \cite{badrinarayanan2017segnet} as the base architecture. We performed segmentation experiments using the following segmentation models, increasingly including the three architectural modifications to the basic SegNet network:
\begin{itemize}
\item[M1:] SegNet model (SN)
\item[M2:] Segnet model with Skip Connections (SC)
\item[M3:] SegNet+SC+Residual Convolutions (RC)
\item[M4:] SegNet+SC+RC+Attention Mechanism (AM)
\end{itemize}

Across all experiments, the loss function chosen is the Focal Loss \cite{lin2017focal}. The segmentation maps often have a class imbalance where the background (black) pixels are more than the foreground (white) pixels. The Focal Loss function handles class imbalance using two factors: the modulating factor and the focusing parameter hence it is chosen as the loss function. The Focal loss function is defined as
\begin{equation}
\text{Focal Loss} = -\alpha_t(1 - p_t)^\gamma \log(p_t)
\label{eq:focal_loss}
\end{equation}
where
\begin{align*}
    \alpha_t & : \text{Weighing factor} \quad (\alpha_t \in [0, 1]) \\
    p_t & : \text{Estimated probability} \quad (p_t \in [0, 1]) \\
    (1 - p_t)^\gamma & : \text{Modulating factor} \\
    \gamma & : \text{Focusing parameter} \quad (\gamma \geq 0)
\end{align*}
The cross-entropy loss function, defined as $-\log(p_t)$, is enhanced by including the Modulating Factor $(1 - p_t)^\gamma$. This modulating factor can be tuned using the Focusing Parameter $\gamma \geq 0$ as shown in the equation \ref{eq:focal_loss}. 
\subsection{Results and Discussion}

\subsubsection{Region-based measures}
The segmentation accuracy is quantitatively measured using IoU, TPR, FNR, TNR, FPR, and Dice Score. Table \ref{table:1} shows the corresponding values of these measures observed on the PH2 dataset. The proposed final model (SN+SC+RC+AM) outperforms the U-Net model by about 15\% and the base SegNet architecture by about 6\% in terms of IoU. Quantitatively, the IoU measure is more penalizing compared to Dice Score (refer Eq. \ref{eq:iou} and Eq. \ref{eq:dice}); hence, the values present in the Dice Score column are lower than the mean (IoU) column. A gradual increase in performance after each inclusion reinforces the choice of a particular component. In measures such as FPR and FNR, the improvement seems numerically lesser, raising suspicion about the model's performance gain's statistical significance. This was verified by employing a non-parametric Wilcoxon rank-sum test for statistical significance, and the distribution of FNR and FPR values between the base model and the proposed IARS SegNet was proved to be statistically significant with a p-value of 0.0007 (more statistical difference between the distributions).  

\begin{table}[!hbt]
\resizebox{\columnwidth}{!}{%
\begin{tabular}{|p{3.5cm}|c|c|c|c|c|c|}
\hline
\text{\bf Model} & \text{\bf TPR} & \text{\bf FPR} & \text{\bf TNR} & \text{\bf FNR} & \text{\bf Dice Score} & \text{\bf mean (IoU)} \\
\hline
\text{U-Net \cite{al2018skin}} & - & - & - & - & 87.61\% & 77.95\% \\
\hline
\text{SN} & 90.23\% & 0.11\% & 94.30\% & 0.09\% & 92.77\% & 86.41\% \\
\hline
SN+SC & 92.5\% & 0.08\% & 95.51\% & 0.07\% & 94.53\% & 88.44\% \\
\hline
SN+SC+RC & 94.22\% & 0.04\% & 96.32\% & 0.02\% & 95.18\% & 91.39\% \\
\hline
SN+SC+RC+AM & \textbf{96.46\%} & \textbf{0.04\%} & \textbf{98.94\%} & \textbf{0.01\%} & \textbf{97.12\%} & \textbf{92.33\%} \\
\hline
\end{tabular}%
}
\caption{Segmentation results for the different model combinations. SN: SegNet, SC: Skip Connections, RC: Residual Convolutions, AM: Attention Mechanism}
\label{table:1}
\end{table}

\begin{figure}[ht]
\centering
\includegraphics[width = \textwidth]{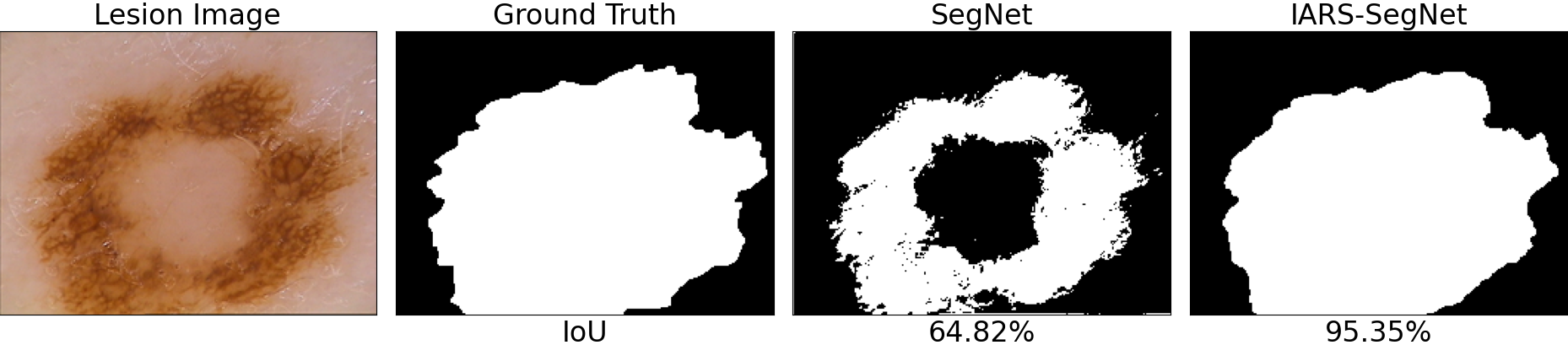}
\caption{Unclear lesion boundary image where inclusion of AM and RC gives better segmentation}
\label{fig:anamoly}
\end{figure}

\begin{figure}[ht]
\centering
\includegraphics[width = \textwidth]{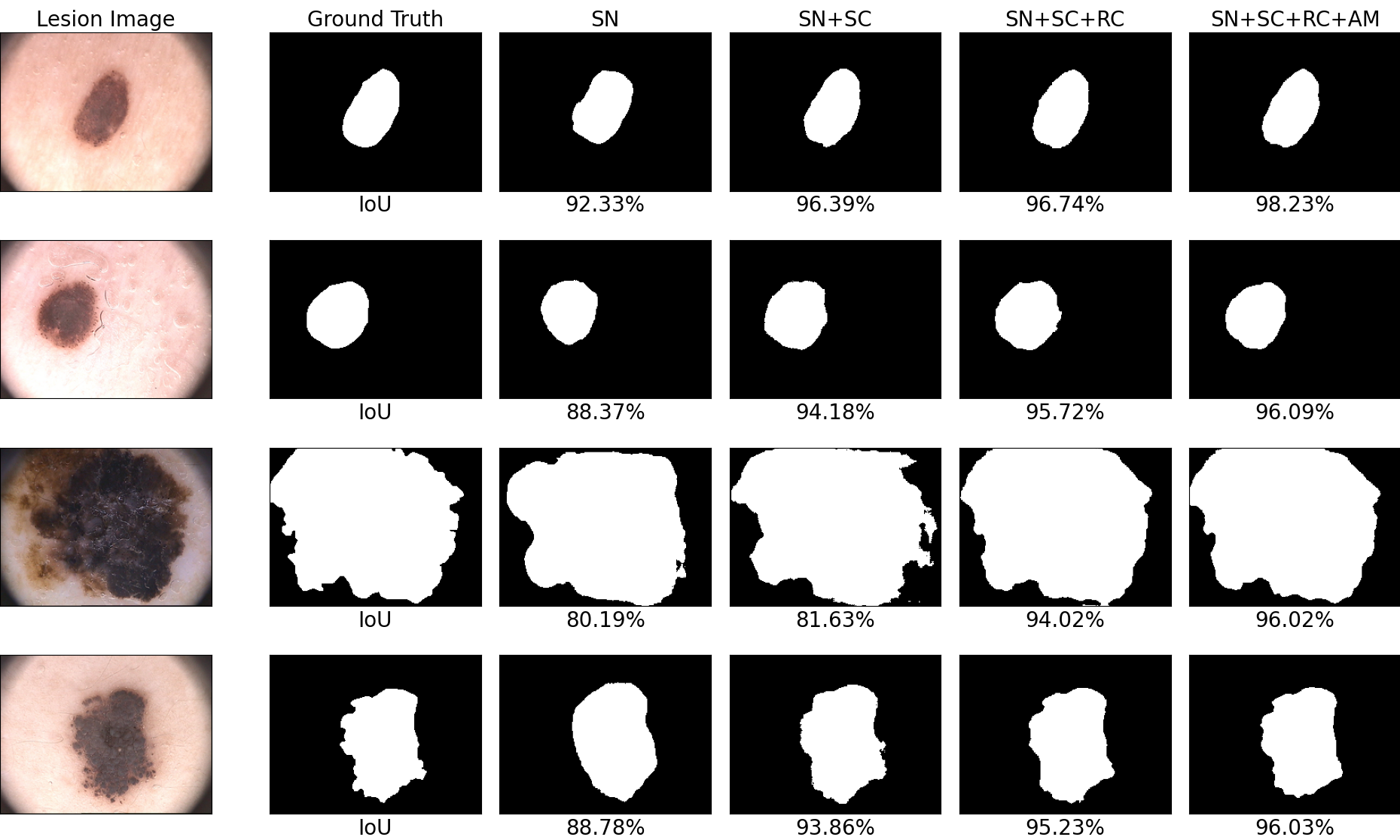}
\caption{Example ground truth and predicted segmentation masks}
\label{iou_panel}
\end{figure}

Figure \ref{iou_panel} showcases some example segmentation maps obtained after the inclusion of each of the components. The segmentation map obtained after the inclusion of the Residual Convolutions and the Attention Mechanism shows superior performance due to its ability to distinguish the unique pigment pattern found along the exterior regions of a lesion. Furthermore, for lesions with fuzzy boundaries or disjoint pigment pattern, figure \ref{fig:anamoly} is one such case where the model without Residual Convolutions and Attention Mechanism fails. This leads to fewer False Positives and more overlap between the segmentation mask and the ground truth.  

\subsubsection{Contour-based measures}

The evaluation of the contour details can be done visually as well as quantitatively. EFD and Hu Moments are used to quantify the contours of the lesion. The closeness of the EFD and Hu Moment vectors of the predicted and ground truth segmentation mask's contour are estimated using the Euclidean distance measure and the Mahalanobis distance measure, respectively. Table \ref{table:1} contains the average distances between the ground truth and predicted segmentation mask's contours across all the images in the PH2 dataset. A distance measure closer to 0 indicates high similarity in the contours. In other words, the model with lower distance measures (both EFD and Hu Moments) implies a better ability to capture the contour details. The proposed IARS SegNet model outperforms the base SegNet architecture in the distance measures represented using EFD and Hu Moments. This indicates that the model is not only able to exhibit superior performance in terms of segmentation accuracy (region-based metrics as shown in the previous section) but is also receptive to the contours of the skin lesion.
\begin{table}[ht]
\centering
\begin{tabular}{|c|c|c|}
\hline
\textbf{Model} & \textbf{EFD} & \textbf{Hu Moments} \\
\hline
SN & 1.44 & 0.35 \\
\hline
SN+SC+RC+AM & 1.01 & 0.30 \\
\hline
\end{tabular}
\caption{Contour performance measures. SN: SegNet, SC: Skip Connections, RC: Residual Convolutions, AM: Attention Mechanism}
\label{table:2}
\end{table}

\begin{figure}[ht]
\centering
\includegraphics[width = \textwidth]{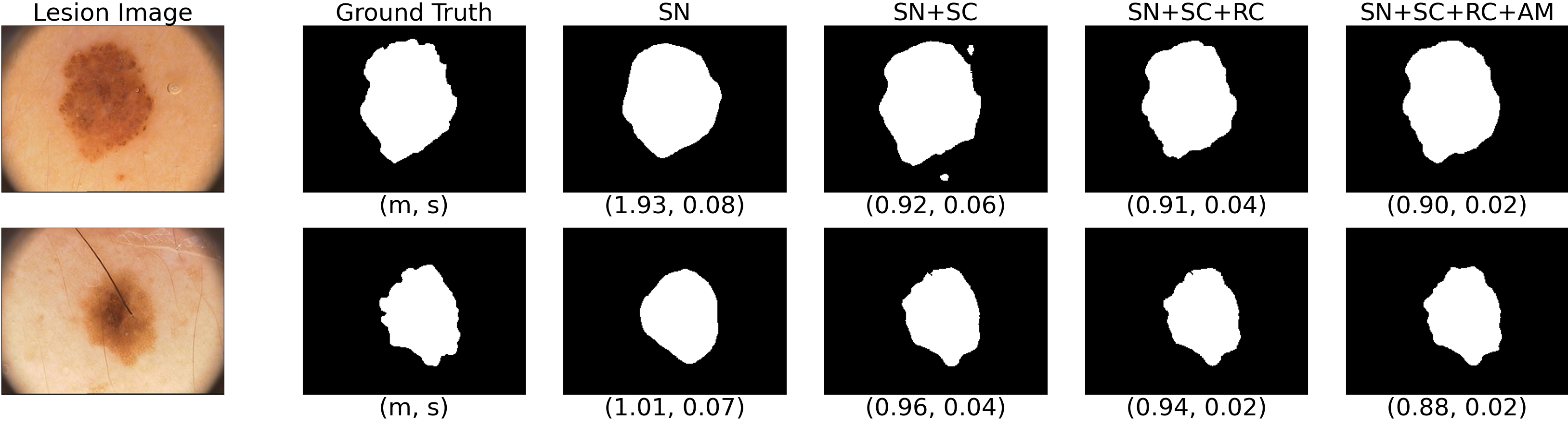}
\caption{Example images with Euclidean (m) and Mahalanobis (s) distance measure values}
\label{contour_panel}
\end{figure}

Figure \ref{contour_panel}  shows the segmentation masks obtained for sample images from the PH2 dataset. The tuple (m, s), represents the Euclidean distance between the EFD vectors and the Mahalanobis distance between the Hu Moments of the ground truth and the consequent model's predicted segmentation masks respectively.  
The segmentation mask obtained after including the Skip Connection is marginally better than the original SegNet architecture. This is due to the direct concatenation of features to enhance gradient flow in the network. It is evident that the contour details are captured better after including the Residual Convolution. This could be attributed to the shortcut connections in Residual Convolutions, which enable better gradient flow by adding the input to the transformed output. This forces the network to perform better along the contours to lower the loss. The performance is even better after including the Attention Mechanism, which weights important regions and ignores irrelevant noisy regions (hair, oil bubble, etc.)  

\subsubsection{Interpretability}
\begin{figure}[ht]
\centering
\includegraphics[width = \textwidth]{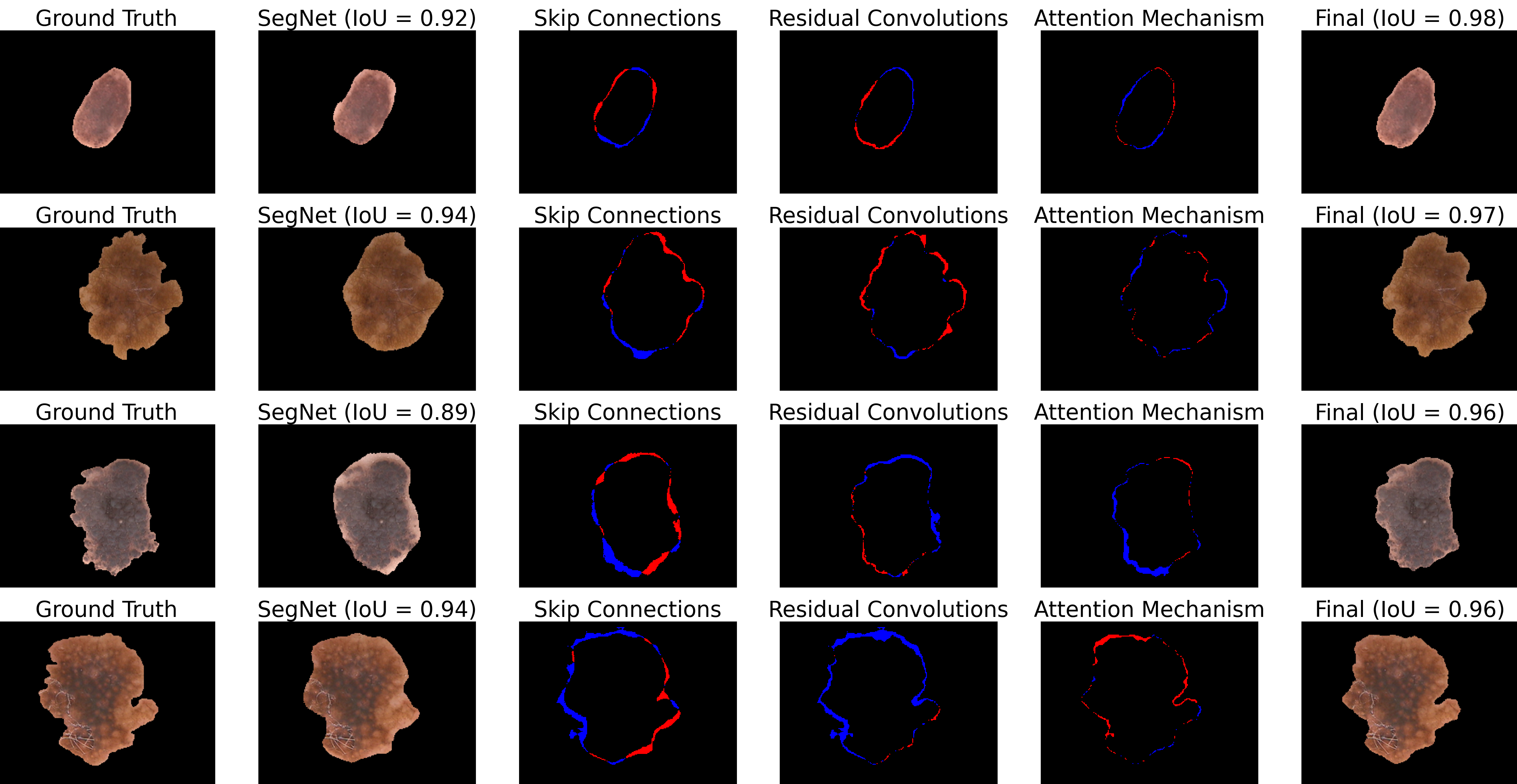}
\caption{Interpretation panel in terms of boundary inclusion/exclusion for each addition of modules on the base Segnet architecture}
\label{interpanel}
\end{figure}

The key feature of the model is its easily interpretable nature. Every modification introduced to the model carries substantial importance and collectively enhances the portrayal of clinically relevant features in the resulting segmentation maps. These inclusions play a pivotal role in refining and enhancing previously predicted segmentation maps by adding or removing specific components. In Figure \ref{interpanel}, we observe the predictions made by the base model and the correction process. This process progressively refines the segmentation, resulting in a more precise and accurate boundary representation.
 The initial segmentation mask from the SegNet is coarse and has smooth contours regardless of the lesion. Then, each column corresponds to the changes introduced to the architecture. Blue represents the pixels included, and red represents those removed by the respective inclusions to the base model. We better understand model improvement by analyzing the $2^{nd}$ and the $6^{th}$ column images. The final model tends to remove almost all the False Positive regions and fill the False Negative regions for better representation of the contours of the ground truth image.  
In the $3^{rd}$, $4^{th}$, and $5^{th}$ columns, the blue and red regions are the additions and the removals made by the corresponding inclusion on the model. The SegNet provides an approximate segmentation of the lesion, which acts as a starting point that gets refined by each inclusion, all with the intent of getting a better representation of the clinically relevant features, thus resulting in superior performance. The visual representation shows how the individual components contribute to the final architecture and thus gives insights into the evolution of the segmentation process upon each inclusion.


\begin{figure}
    \centering

    \begin{subfigure}{0.19\textwidth}
        \includegraphics[width=\linewidth]{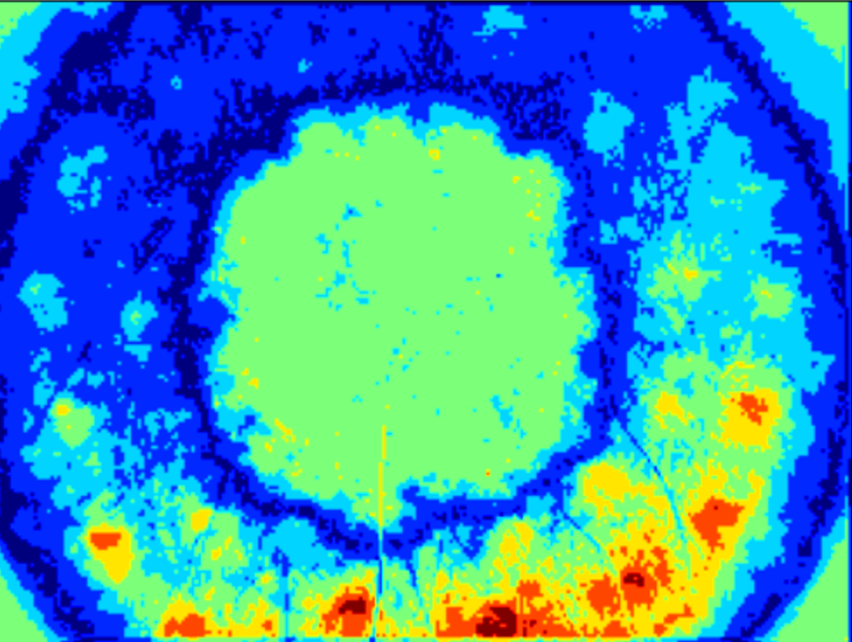}
        \caption*{Convolutional \\block 1}
    \end{subfigure}
    \begin{subfigure}{0.19\textwidth}
        \includegraphics[width=\linewidth]{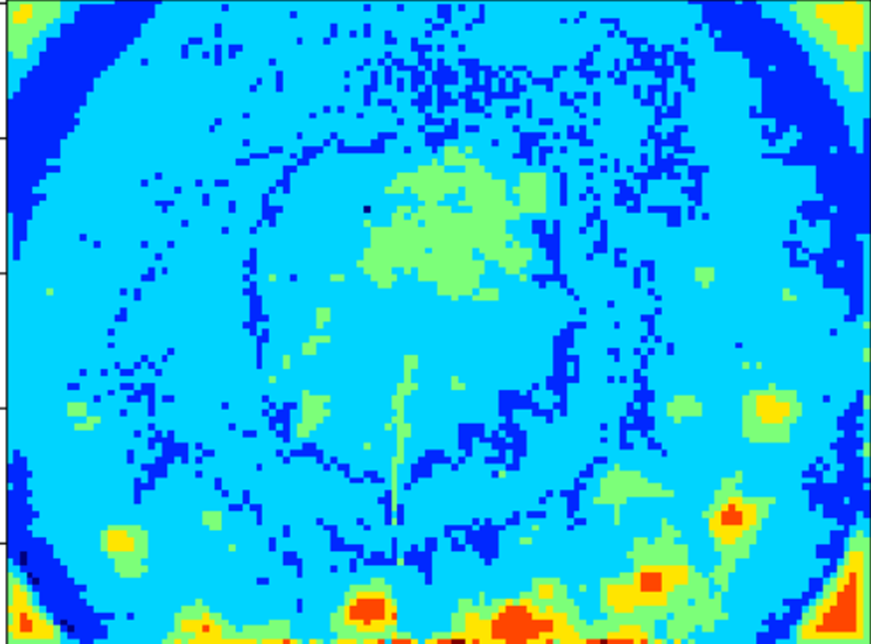}
        \caption*{Convolutional \\block 2}
    \end{subfigure}
    \begin{subfigure}{0.19\textwidth}
        \includegraphics[width=\linewidth]{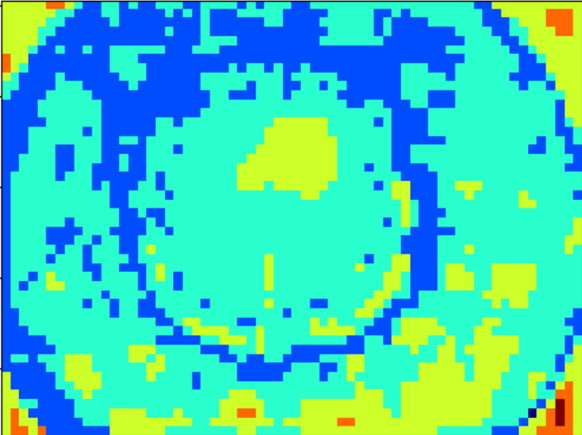}
        \caption*{Convolutional \\block 3}
    \end{subfigure}
    \begin{subfigure}{0.19\textwidth}
        \includegraphics[width=\linewidth]{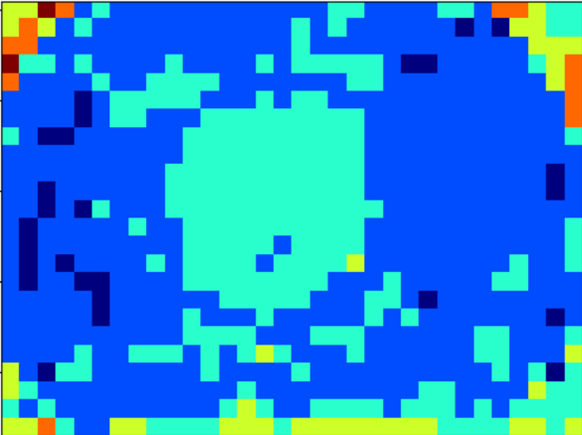}
        \caption*{Convolutional \\block 4}
    \end{subfigure}
    \begin{subfigure}{0.19\textwidth}
        \includegraphics[width=\linewidth]{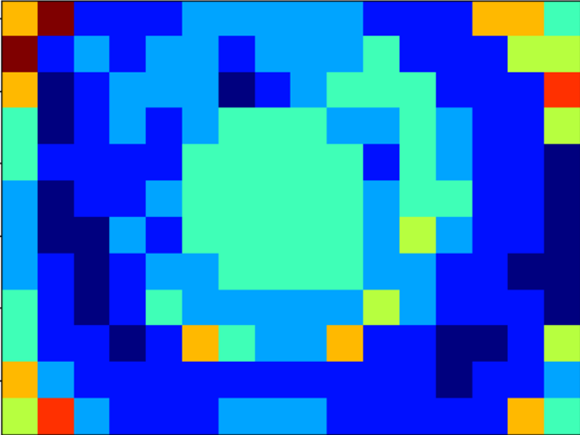}
        \caption*{Convolutional \\block 5}
    \end{subfigure}

    \begin{subfigure}{0.19\textwidth}
        \includegraphics[width=\linewidth]{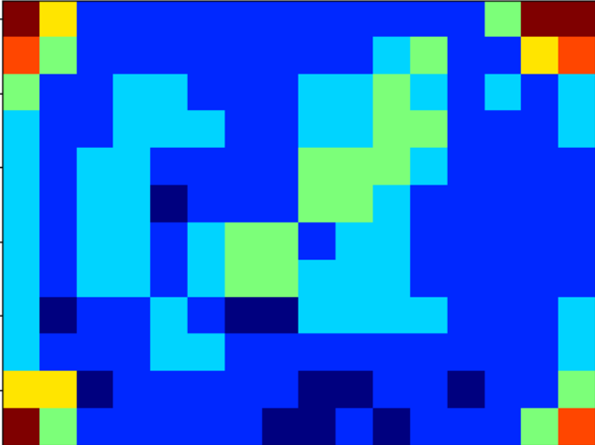}
        \caption*{Deconvolutional \\block 1}
    \end{subfigure}
    \begin{subfigure}{0.19\textwidth}
        \includegraphics[width=\linewidth]{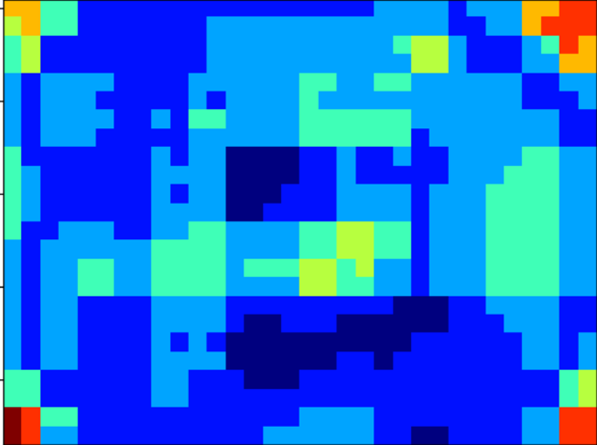}
        \caption*{Deconvolutional \\block 2}
    \end{subfigure}
    \begin{subfigure}{0.19\textwidth}
        \includegraphics[width=\linewidth]{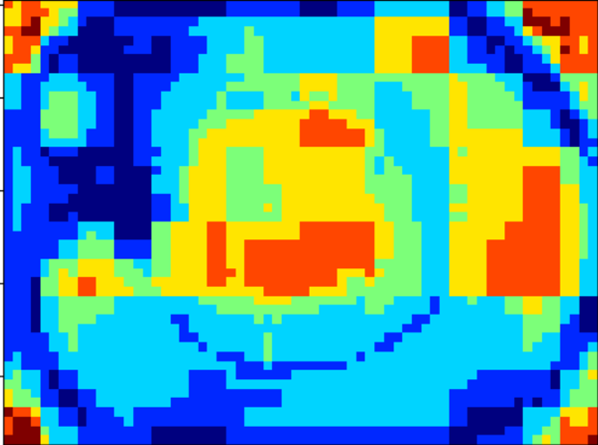}
        \caption*{Deconvolutional \\block 3}
    \end{subfigure}
    \begin{subfigure}{0.19\textwidth}
        \includegraphics[width=\linewidth]{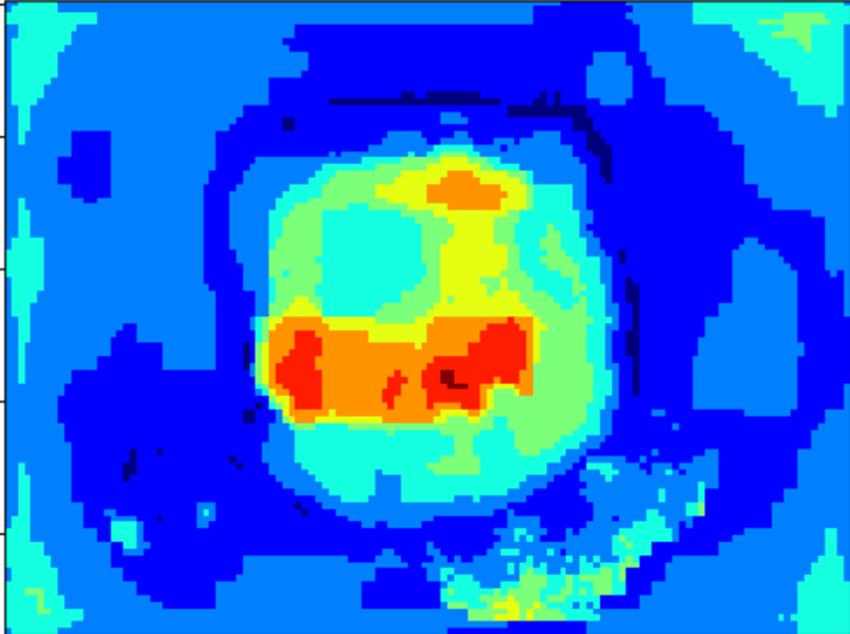}
        \caption*{Deconvolutional \\block 4}
    \end{subfigure}
    \begin{subfigure}{0.19\textwidth}
        \includegraphics[width=\linewidth]{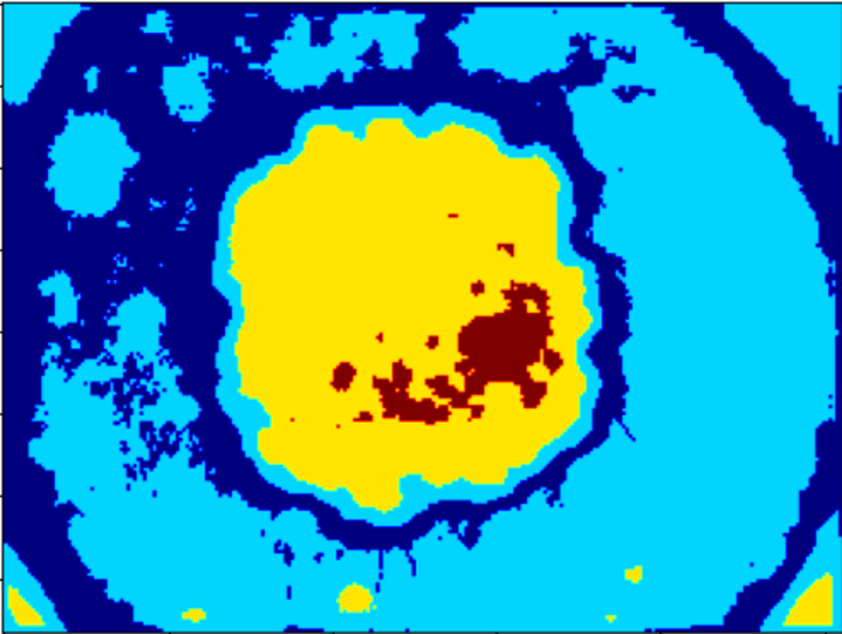}
        \caption*{Deconvolutional \\block 5}
    \end{subfigure}

    \caption{Maximum Intensity Projections}
    \label{MIP}
\end{figure}

To further gain a deeper understanding of how the proposed segmentation model evolves in its localization of the region of interest, Maximum Intensity Projections (MIP) of feature maps from the convolutional/ deconvolutional block are visualized. 
All the images present in Figure \ref{MIP}  consist of MIPs extracted from both the convolutional and deconvolutional blocks within the model. The first row displays MIPs from the convolutional blocks, while the second row showcases MIPs from the deconvolutional blocks. These feature map extractions offer a glimpse into how the model comprehends the semantics of the segmentation task. They reveal how the model effectively segments relevant information, guided by mechanisms such as skip connections, residual convolutions, and attention. These feature maps also serve as heat maps, highlighting areas of focus for the model. These visualizations provide a more profound insight into the actual segmentation process, shedding light on the underpinnings of the segmentation task. For instance, the segmentation process should not be completely based on the color disparity between the lesion and the background skin. Such a model will be no better than a thresholding algorithm. The proposed model utilizes several clinically relevant features such as lesion boundary and color to drive the segmentation process. 

\section{Conclusion}\label{sec:conclusion}

In this paper, we presented an enhanced SegNet architecture tailored for skin lesion segmentation and conducted a thorough performance analysis using region-based and contour-based evaluation metrics to underscore its significance in melanoma segmentation. Our approach involved the integration of three pivotal components: Skip Connections (SC), Residual Convolutions (RC), and Attention Mechanisms (AM), which exhibited a progressive enhancement in the model's performance and its ability to extract clinically relevant features. Furthermore, we substantiated each of these inclusions by grounding them in the features extracted from skin lesions and their clinical relevance to melanoma classification. To gauge the effectiveness of these components, we compared the proposed model with intermediate models created after the incorporation of each element, employing region and contour-based evaluation metrics. The results decisively highlighted a comprehensive improvement in the model's performance, affirming the judicious inclusion of each component. To bolster the model's interpretability, we provided two types of visualizations: Maximum Intensity Projections from each convolutional and deconvolutional block, as well as the regions added and removed by introducing these components to the base SegNet model. These visual aids empower physicians to place their trust in the segmentation maps generated by this AI system, rendering the proposed architecture a practical choice for real-world clinical applications. Our experiments culminated in the finding that the proposed architecture outperforms both the SegNet and U-Net models in the task of skin lesion segmentation. Looking ahead, future research could explore the incorporation of additional clinically relevant skin lesion features, such as texture, into the architecture to enhance its robustness and adaptability.\\\\
\textbf{Acknowledgements}\\\\This research was funded by the Spanish Ministry of Science and Innovation, grant number PID2020-116927RB-C22 (R.B.).

\appendix

\section{Hu Moments derivation}\label{secA1}
The seven Hu Moments can be derived through the following equations, 
\begin{equation}
         m_{pq} = \int_{-\infty}^{\infty} \int_{-\infty}^{\infty} x^p y^q f(x,y) \, dx \, dy
         \label{eq:moments}
    \end{equation}
    where $p,q = 0, 1, 2, \ldots$

    Equation \ref{eq:moments} represents the $(p+q)^{th}$ order moments. These moments are not invariant to translation, rotation, and scale. 

    \begin{equation}
    \mu_{pq} = \int_{-\infty}^{\infty} \int_{-\infty}^{\infty} (x - \bar{x})^p (y - \bar{y})^q f(x,y) \, dx \, dy
    \label{eq:centralmoments}
    \end{equation}
    where $p,q = 0, 1, 2, \ldots$

    The Central moments are calculated as shown in the equation \ref{eq:centralmoments}. These moments are location invariant as it is obtained by shifting the moments to the centroid of the image $f(x,y)$
    
    \begin{equation}
            \bar{x} = \frac{m_{10}}{m_{00}} 
            \label{eq:xcentroid}
    \end{equation}
    \begin{equation}
            \bar{y} = \frac{m_{01}}{m_{00}}
            \label{eq:ycentroid}
    \end{equation}
    The centroid of the image $f(x,y)$ can be calculated from the equations \ref{eq:xcentroid} and \ref{eq:ycentroid}. Centering the moments from equation \ref{eq:moments} to $(\bar{x}, \bar{y})$  yields the central moments, as formulated in equation \ref{eq:centralmoments}.
    \begin{equation}
    \eta_{p q}=\frac{\mu_{p q}}{\mu_{00}^\gamma}, \quad \gamma=(p+q+2) / 2
    \label{eq:hunormalization}
    \end{equation}
    where $p+q = 2, 3, \ldots$ 
    
    Furthermore, the scale invariance can be achieved by normalizing the central moments. The normalized central moments can be obtained by dividing the central moments by the $0$ order moments raised to the power of $\gamma$ as shown in equation \ref{eq:hunormalization}. Using these normalized central moments, Hu Moments are described as follows
\begin{align*}
\begin{split}
\phi_1 &= \eta_{20}+\eta_{02}
\end{split}
\end{align*}

\begin{align*}
\begin{split}
\phi_2 &= \left(\eta_{20}-\eta_{02}\right)^2+4 \eta_{11}^2
\end{split}
\end{align*}

\begin{align*}
\begin{split}
\phi_3 &= \left(\eta_{30}-3 \eta_{12}\right)^2+\left(3 \eta_{21}-\mu_{03}\right)^2
\end{split}
\end{align*}

\begin{align*}
\begin{split}
\phi_4 &= \left(\eta_{30}+\eta_{12}\right)^2+\left(\eta_{21}+\mu_{03}\right)^2
\end{split}
\end{align*}

\begin{align*}
\begin{split}
\phi_5 &= \left(\eta_{30}-3 \eta_{12}\right)\left(\eta_{30}+\eta_{12}\right)\left[\left(\eta_{30}+\eta_{12}\right)^2-3\left(\eta_{21}+\eta_{03}\right)^2\right] \\
&\quad+\left(3 \eta_{21}-\eta_{03}\right)\left(\eta_{21}+\eta_{03}\left[3\left(\eta_{30}+\eta_{12}\right)^2-\left(\eta_{21}+\eta_{03}\right)^2\right]\right)
\end{split}
\end{align*}

\begin{align*}
\begin{split}
\phi_6 &= \left(\eta_{20}-\eta_{02}\right)\left[\left(\eta_{30}+\eta_{12}\right)^2-\left(\eta_{21}+\eta_{03}\right)^2\right] \\
&\quad+4 \eta_{11}\left(\eta_{30}+\eta_{12}\right)\left(\eta_{21}+\eta_{03}\right)
\end{split}
\end{align*}

\begin{align*}
\begin{split}
\phi_7 &= \left(3 \eta_{21}-\eta_{03}\right)\left(\eta_{30}+\eta_{12}\right)\left[\left(\eta_{30}+\eta_{12}\right)^2-3\left(\eta_{21}+\eta_{03}\right)^2\right] \\
&\quad-\left(\eta_{30}-3 \eta_{12}\right)\left(\eta_{21}+\eta_{03}\right)\left[\left(3\left(\eta_{30}+\eta_{12}\right)^2-\left(\eta_{21}+\eta_{03}\right)^2\right]\right)
\end{split}
\end{align*}




 \bibliographystyle{elsarticle-num} 
 \bibliography{cas-refs}





\end{document}